\documentclass{article}
\usepackage{spconf,amsmath,graphicx}
\usepackage{amssymb}
\usepackage{booktabs}
\usepackage{multirow}
\usepackage{graphicx}

\usepackage{soul}
\usepackage{color}
\usepackage[table,xcdraw]{xcolor}

\usepackage{enumitem}
\setlist{nosep, leftmargin=14pt}
\usepackage{mwe} 

\definecolor{electricindigo}{rgb}{0.44, 0.0, 1.0}

\title{ViGU: Vision GNN U-Net for fast MRI}

\name{Jiahao Huang$^{1,2,\star}$,
\quad Angelica I. Aviles-Rivero$^{3}$,
\quad Carola-Bibiane Sch{\"o}nlieb$^{3}$, 
\quad Guang Yang$^{1,2,\star}$}

\address{
$^{1}$ National Heart and Lung Institute, Imperial College London, United Kingdom 
\and $^{2}$ Cardiovascular Research Centre, Royal Brompton Hospital, United Kingdom 
\and $^{3}$ Department of Applied Mathematics and Theoretical Physics, University of Cambridge, United Kingdom \\~\\ 
$^{\star}$ Send correspondence to \{j.huang21,g.yang\}@imperial.ac.uk}

\begin{document}
%
\maketitle
\begin{abstract}
Deep learning models have been widely applied for fast MRI. The majority of existing deep learning models, e.g., convolutional neural networks, work on data with Euclidean or regular grids structures. However, high-dimensional features extracted from MR data could be encapsulated in non-Euclidean manifolds. This disparity between the go-to assumption of existing models and data requirements limits the flexibility to capture irregular anatomical features in MR data. In this work, we introduce a novel Vision GNN type network for fast MRI called Vision GNN U-Net (ViGU). More precisely, the pixel array is first embedded into patches and then converted into a graph. Secondly, a U-shape network is developed using several graph blocks in symmetrical encoder and decoder paths. Moreover, we show that the proposed ViGU can also benefit from Generative Adversarial Networks yielding to its variant ViGU-GAN. We demonstrate, through numerical and visual experiments, that the proposed ViGU and GAN variant outperform existing CNN and GAN-based methods. Moreover, we show that the proposed network readily competes with approaches based on Transformers while requiring a fraction of the computational cost. More importantly, the graph structure of the network reveals how the network extracts features from MR images, providing intuitive explainability.
\end{abstract}


\section{Introduction}
\label{sec:intro}

Magnetic Resonance Imaging (MRI) is one of the most important clinical tools. It provides high-resolution and non-invasive imaging for diagnosis and prognosis in a harmless manner. However, MRI has an inherently slow scanning time, since the raw data is acquired in \textit{k}-space, and the minimum scanning time is decided by the selection of temporal and spatial resolution as well as the field of view, constraining by the Nyquist theorem. The prolonged scanning time leads to artefacts from the voluntary and involuntary physiological movements of the patients~\cite{Chen2022ai}.

With the thriving development of artificial intelligence technologies, deep learning-based models have been promptly developed for fast MRI~\cite{Yang2016admm,Schlemper2018dccnn}. 
Convolutional neural networks (CNNs) have dominated research studies in computer vision (CV) and medical image analysis, including MRI reconstruction~\cite{Schlemper2018dccnn, Yang2018dagan, Huang2021piddgan}, taking advantage of the inductive biases of locality and weight sharing, and their hierarchical structures. 
Recently, Transformers~\cite{Dosovitskiy2020vit} have shown superiority for CV tasks bolstered by their global sensitivity and long-range dependency. Transformer-based MRI reconstruction methods~\cite{KorkmazDeep2021,Huang2022swinmr,Huang2022stgan,Huang2022sdaut} have been proposed and achieved promising results, even though their increased computational cost is still a challenge for a wider application.
\begin{figure}[t!]
    \centering
    \centerline{\includegraphics[width=3.5in]{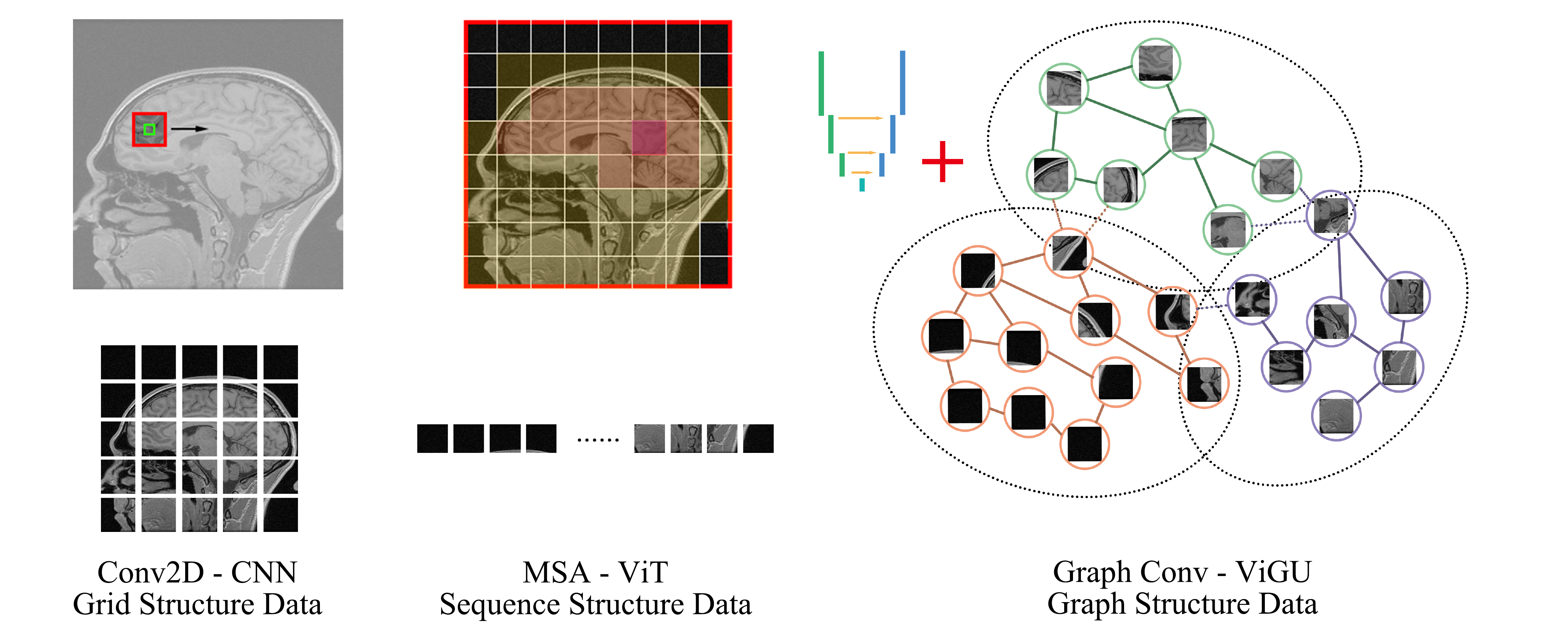}}
  \caption{Data Structures of Convolutional Neural Networks, Vision Transformers and Vision GNN U-Net.}
    \label{fig:FIG_Overview}
\end{figure}

General CNNs and Transformers backbones treat image data differently (Figure~\ref{fig:FIG_Overview}). The 2D convolution (Conv2D) in CNNs applies sliding operation kernel on pixels in a regular grid, exploiting the shift-invariance and local prior.
The multi-head self-attention (MSA) in Transformers (specifically ViT~\cite{Dosovitskiy2020vit}), embeds different ranges of pixels into patches, then converts them into sequences, introducing global sensitivity and long-range dependency. However, both Conv2D and MSA operations are usually based on the regular pixel grid in the Euclidean space~\cite{Han2022vig}.

Recently, Han et al. in~\cite{Han2022vig} proposed the Vision GNN (ViG) backbone. ViG combines, combining the patch embedding from ViT~\cite{Dosovitskiy2020vit} and the idea of Graph Convolutional Networks (GCNs)~\cite{Li2019gcn}, treating images with more flexibility from the graph perspective. 
GCNs are originally designed for tackling specific tasks for non-Euclidean data, e.g., point cloud, social network, and biochemical graphs. Vision GNN fills the technological gap between GNNs and image data for computer vision tasks, and achieves state-of-the-art results in high-level tasks like classification and detection tasks.

For MR images, the shape of anatomical structures are irregular, leading to redundancy and inflexibility when using the conventional grid or sequence data structure. 
We hypothesise that treat MR images as graphs (Figure~\ref{fig:FIG_Overview}) can provide a comprehensive understanding of the anatomical structures in MR images.
Specifically, the image is first converted into patches by a shallow CNN and then regarded as nodes in a graph. 
Nodes with similar features can be gathered and connected using the K-nearest neighbours (KNN) algorithm, where information exchange can be conducted.
Different anatomical structures can be recognised as sub-graphs of the whole graph (for an image). The edge connections within and between sub-graphs can be learnt to reflect the intra- and inter-relationship of anatomical structures.

In this paper, we exploit how ViG works for a specific low-level image restoration task, i.e., MR reconstruction, by introducing a ViG-based U-Net, namely ViGU, and its variants based on Generative Adversarial Network (GAN), namely ViGU-GAN. Experiments have shown that our proposed ViGU and ViGU-GAN can outperform CNN-based and GAN-based MRI reconstruction methods and can achieve comparable results with Transformer-based methods with much a lower computational cost. The edge connection of ViGU shows that the proposed ViGU can learn the intra- and inter-relationship of different anatomical structures, providing model explainability.

\section{Methods}\label{sec:method}
This section describes in detail the key parts of the proposed ViGU network and variant.

\subsection{U-Net Based Architecture}
\begin{figure}[thpb]
    \centering
    \centerline{\includegraphics[width=3.2in]{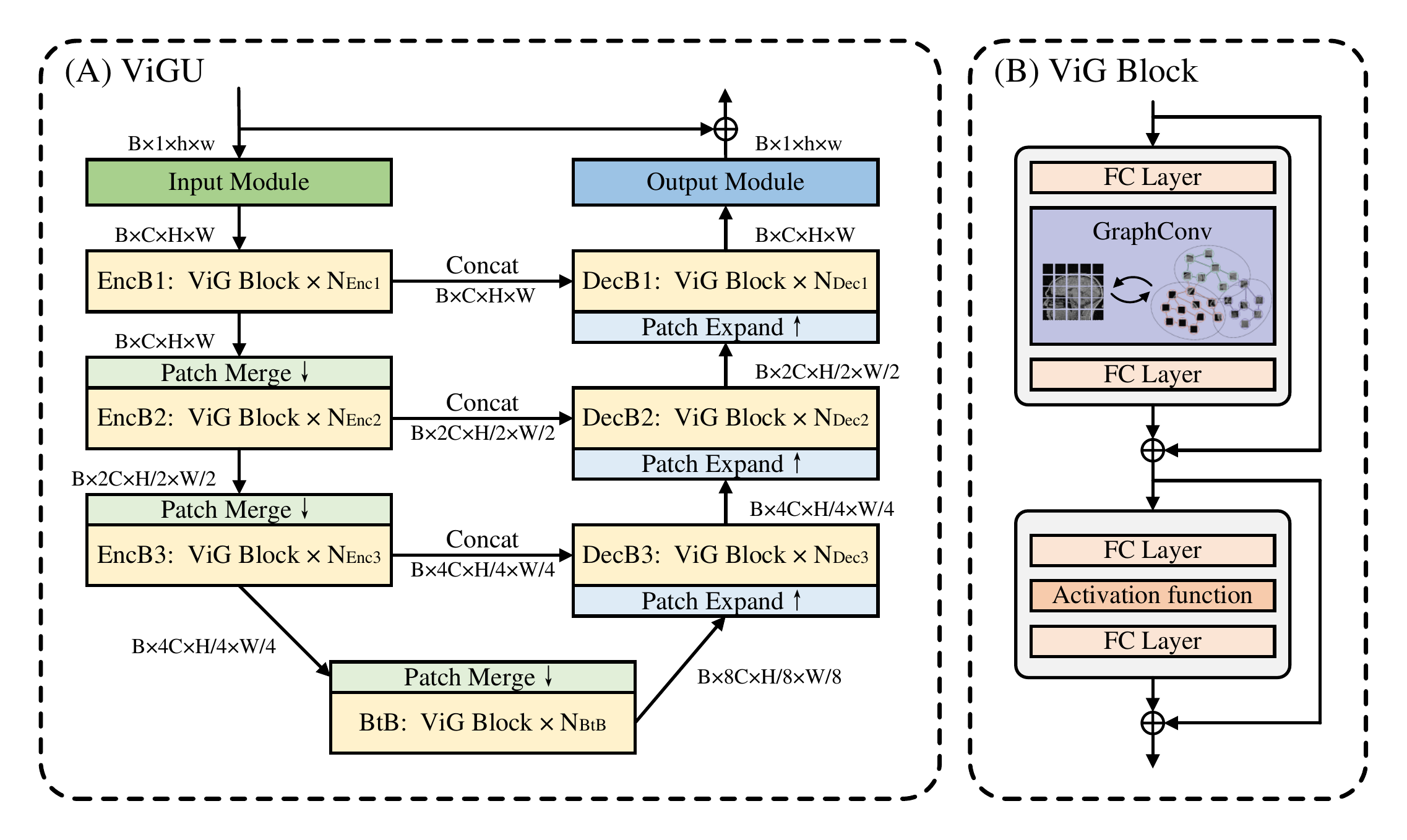}}
    \caption{(A) The network architecture of our ViGU; (B) The structure of the ViG Blocks.}
    \label{fig:FIG_Structure}
\end{figure}

The architecture of the proposed ViGU is displayed in Fig.~\ref{fig:FIG_Structure} (A).
CNN-based input and output modules are applied at the beginning and end of our ViGU converting between images $\mathbb{R}^{h \times w \times 1}$ and patch vectors $\mathbb{R}^{N \times C}$. We denote $r$ and $C$ as the patch size and embedding channel number respectively. We then define the number of patches as $N = H \times W = h/r \times w/r$. Relative position embedding is applied for each patch, which is omitted for brevity. 

Three encoder blocks (EncB) and three decoder blocks (DecB) are symmetrically arranged in the encoder and decoder path correspondingly, between which a bottleneck block (BnB) is placed. 
The EncB, DecB and BnB are composed of one or multiple ViG Blocks, which are the basic computation blocks for ViGU.
The resolution of feature maps is gradually decreased and increased along the encoder and decoder paths. Information is passed, via the skip connection and concatenation operation, from the encoder to the decoder paths between feature maps with the same resolution. Residual connection is applied to convert the ViGU into a refinement function: $\hat x_u = \text{H}_\text{ViGU}(x_u) + x_u$.

\subsection{Graph-level Operation}
A key step is how to transform an image as a graph $\mathcal{G}=(\mathcal{V}, \mathcal{E})$ composed of a set of of nodes $\mathcal{V}$ connected by a set of edges $\mathcal{E}$.
For each feature map, we have a group of patches $X = \{x_1, x_2, ..., x_N\}$, which are viewed as a set of unordered nodes $\mathcal{V} = \{v_1, v_2, ..., v_N\}$. For a single node $v_i$, $K$ edges $\mathcal{E}_i = \{e_{1i}, e_{2i}, ..., e_{Ki}\}$ are acquired from its $K$ nearest neighbours $\mathcal{N}(v_i)$, where $e_{ji}$ indicates the edge from node $v_j$ to node $v_i$. 

The graph representation of feature map $X$ can be expressed as $\mathcal{G}(X)$. 
A graph convolution operation $\text{H}_\text{GConv}$ is expressed as:
\begin{align}\label{eq:gc}
\mathcal{G}^{\prime}
&= \text{H}_\text{GConv}(\mathcal{G}(X), W) \nonumber \\
&= \text{H}_\text{Update}(\text{H}_\text{Aggregate}(\mathcal{G}(X), W_\text{Aggregate}), W_\text{Update}),
\end{align}
\noindent in which $\text{H}_\text{Aggregate}$ and $\text{H}_\text{Update}$ refer to the Aggregate and Update operations in graph convolution with learnable parameters $W_\text{Aggregate}$ and $W_\text{Update}$~\cite{Han2022vig}.

\subsection{ViG Block}
As Figure~\ref{fig:FIG_Structure} (B) shows, ViG Block adopted the structure from ViT Block~\cite{Dosovitskiy2020vit}, which can be expressed as:
\begin{align}\label{eq:vig_block}
& X^{\prime} = \text{FC}(\text{GraphConv}(\text{FC}(X))) + X \\
& X^{\prime\prime} = \text{MLP}(X) + X^{\prime},
\end{align}
\noindent where $X$ and $X^{\prime\prime}$ are the input and output of ViG Block. $\text{GraphConv}(\cdot)$ and $\text{MLP}(\cdot)$ denote the graph convolution and the multi-layer perceptron. $\text{FC}(\cdot)$ denotes the full connected layer, which is applied before and after the graph convolution; with the purpose to keep the domain consistency between node and image features and increase the feature diversity. All the normalisation and activation functions are omitted for brevity.

\subsection{Optimisation Scheme}
To train our proposed ViGU, Charbonnier loss is applied to the image and frequency domains, which are denoted as $\mathcal{L}_{\mathrm{img}}(\theta)$ and  $\mathcal{L}_{\mathrm{freq}}(\theta)$ respectively. They allow for constraining the ground truth MR images $x$ and reconstructed MR images $\hat x_u$. Moreover, a $l1$ loss is applied for perceptual-based, $\mathcal{L}_{\mathrm{perc}}(\theta)$, constraints using a pre-trained VGG $f_{\mathrm{VGG}}(\cdot)$. Formally, they read:
\begin{align}\label{eq:loss_list}
&\mathop{\text{min}}\limits_{\theta} 
\mathcal{L}_{\mathrm{img}}(\theta) =
\sqrt{\mid\mid x - \hat x_u \mid\mid^2_2 + \epsilon^2},\\
&\mathop{\text{min}}\limits_{\theta} 
\mathcal{L}_{\mathrm{freq}}(\theta) =
\sqrt{\mid\mid \mathcal{F}x - \mathcal{F} \hat x_u \mid\mid^2_2 + \epsilon^2},\\
&\mathop{\text{min}}\limits_{\theta} 
\mathcal{L}_{\mathrm{perc}}(\theta) =
\mid\mid f_{\mathrm{VGG}}(x) - f_{\mathrm{VGG}}(\hat x_u) \mid\mid_1,
\end{align}
\noindent where $\epsilon$ is empirically set to $10^{-9}$. We denote $\theta$ as the network parameter of ViGU, and $\mathcal{F}$ refers to the Fourier transform. 
The total loss of ViGU, $\mathcal{L}_{\mathrm{ViGU}}(\theta)$, using is computed as:
\begin{align}\label{eq:loss_total}
\mathcal{L}_{\mathrm{ViGU}}(\theta)
= \alpha \mathcal{L}_{\mathrm{img}}(\theta)
+ \beta \mathcal{L}_{\mathrm{freq}}(\theta)
+ \gamma \mathcal{L}_{\mathrm{perc}}(\theta),
\end{align}
\noindent where $\alpha$, $\beta$ and $\gamma$ are weighting parameters balancing the importance of each term.

Our ViGU can also benefit from GAN principles yielding to a new variant called ViGU-GAN. For the GAN-based variant, the proposed ViGU is the generator $G_{\theta_{G}}$ parameterised by $\theta_{G}$ (same with the $\theta$ in ViGU), and a U-Net based discriminator~\cite{Schonfeld2020}, $D_{\theta_{D}}$, is applied for adversarial training. The adversarial loss $\mathcal{L}_{\mathrm{adv}}(\theta_{G}, \theta_{D})$ is then given by:
\begin{align}\label{eq:loss_adv}
&\mathop{\text{min}}\limits_{\theta_{G}} 
\mathop{\text{max}}\limits_{\theta_{D}}
\mathcal{L}(\theta_{G}, \theta_{D}) \\
&=\mathbb{E}_{x \sim p_{\mathrm{t}}(x)}
[\mathop{\text{log}} D_{\theta_{D}}(x)]
- \mathbb{E}_{x_u \sim p_{\mathrm{u}}(x_u)}
[\mathop{\text{log}} D_{\theta_{D}}(\hat x_u)]. \nonumber
\end{align}
The total loss of ViGU-GAN, $\mathcal{L}_{\mathrm{ViGU-GAN}}(\theta)$, reads:
\begin{align}\label{eq:loss_gan_total}
\mathcal{L}_{\mathrm{ViGU-GAN}}(\theta_{G}, \theta_{D})
= \mathcal{L}_{\mathrm{ViGU}}(\theta_{G})
+ \mathcal{L}(\theta_{G}, \theta_{D}).
\end{align}

\section{Experimental Settings and Results}\label{sec:experiment}
This section describes in detail the set of experiments conducted to validate the proposed ViGU and variant.

\subsection{Implementation Details}
We evaluate our approach using the Calgary-Campinas Public Dataset~\cite{Souza2018}. It is composed of 67 cases of T1-weight 3D brains, and randomly divided into training, validation and testing datasets following a ratio of 6:1:3. The multi-channel data was converted into single-channel MR images using the root sum square method. The top and bottom slices in each case were discarded, and the rest of the 100 slices were chosen for experiments. 

The number of ViG Blocks and embedding channels was set to $[3,3,3,1,3,3,3]$ and $[96,192,384,768,384,192,96]$ respectively. ViGU$_x$ indicated the proposed ViGU with a patch size of $x$. The initial learning rate was set to $6 \times 10^{-4}$ and decays every 10,000 steps by 0.5 from the 50,000$^{\text{th}}$ step. The weighting parameters in the loss function $\alpha$, $\beta$ and $\gamma$ were set to 15, 0.1 and 0.0025. For training the ViGU-GAN, the parameter of the discriminator is updated every 5 steps, to prevent training an ``overly strong'' discriminator and compromising the training of the generator.

We compared the proposed ViGU and ViGU-GAN against MRI reconstruction methods of DAGAN~\cite{Yang2018dagan}, nPIDD-GAN~\cite{Huang2021piddgan} and SwinMR~\cite{Huang2022swinmr} with Gaussian 1D 30\% (G1D30\%) and radial 10\% (R10\%) masks.

For quantitative results, we use Peak Signal-to-Noise Ratio (PSNR), Structural Similarity Index Measure (SSIM), and Fr\'echet Inception Distance (FID)~\cite{Heusel2017}. Multiply Accumulate Operations (MACs) were utilised to estimate the computational complexity with an input size of $1 \times 256 \times 256$.

\subsection{Comparison Experiments}

\begin{figure}[t!]
    \centering
    \centerline{\includegraphics[width=3in]{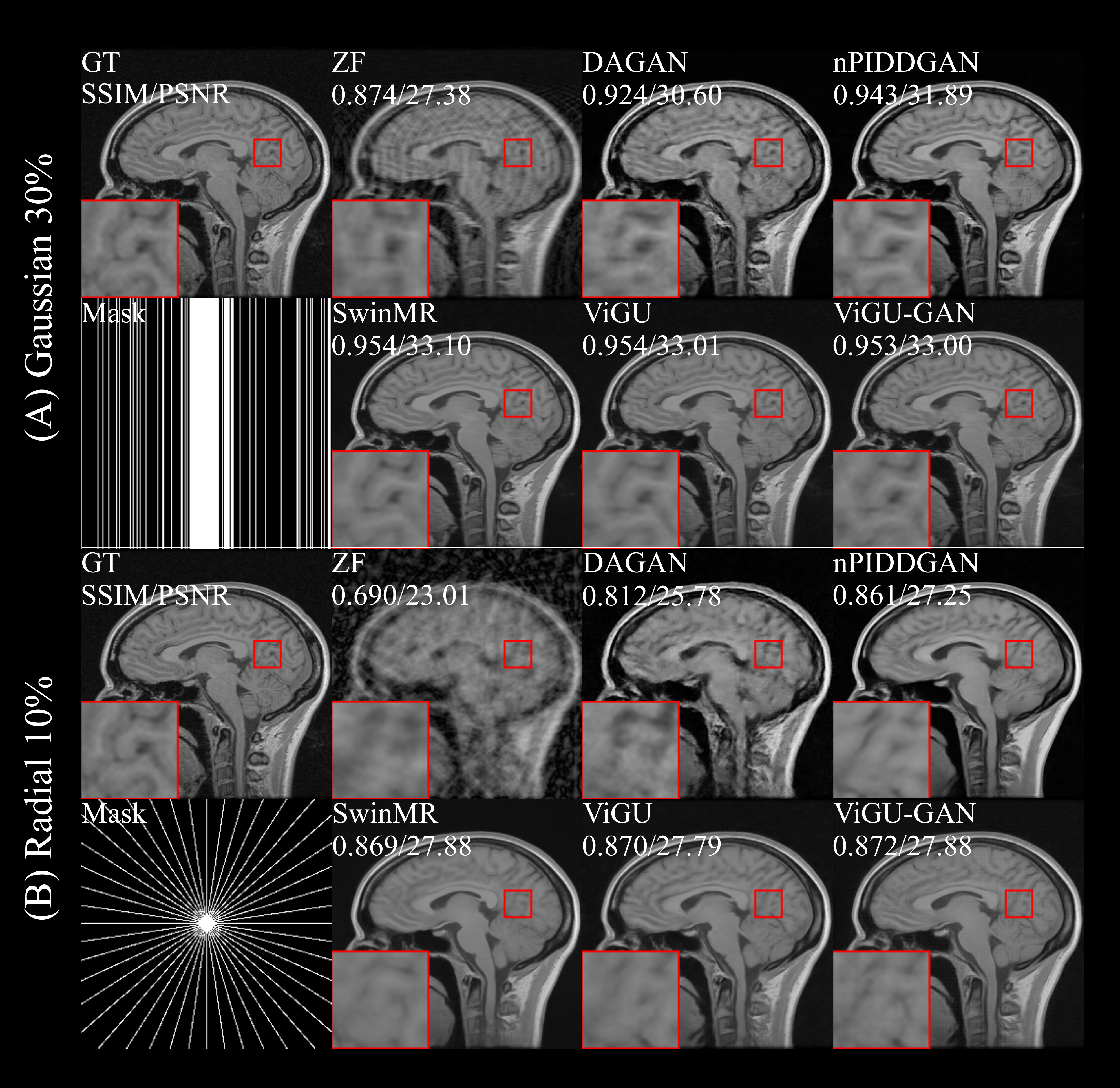}}
    \caption{Visual comparison of our ViGU/ViGU-GAN vs existing techniques. Results display SSIM and PSNR results.}
    \label{fig:FIG_EXP_IMAGE_Comparison}
\end{figure}
\begin{table}[t!]
  \centering
  \caption{Quantitative results of the comparison experiments.}
  \resizebox{\linewidth}{!}{
    \begin{tabular}{cccccccc}
    \toprule
    \multirow{2}[4]{*}{Method} & MACs  & \multicolumn{3}{c}{G1D30\%} & \multicolumn{3}{c}{R10\%} \\
\cmidrule{3-8}          & (G) $\downarrow$   & SSIM $\uparrow$  & PSNR $\uparrow$  & FID $\downarrow$  & SSIM $\uparrow$  & PSNR $\uparrow$  & FID $\downarrow$\\
    \midrule
    ZF    & -     & 0.883 (0.012) & 27.81 (0.82) & 156.38 & 0.706 (0.022) & 23.53 (0.82) & 319.45 \\
    DAGAN & 33.97 & 0.924 (0.010) & 30.41 (0.82) & 56.04 & 0.822 (0.024) & 25.95 (0.85) & 132.58 \\
    nPIDD-GAN & 56.44 & 0.943 (0.009) & 31.81 (0.92) & 26.15 & 0.864 (0.023) & 27.17 (0.97) & 82.86 \\
    SwinMR & 800.73 & \textbf{0.955 (0.009)} & \textbf{33.05 (1.09)} & \underline{21.03} & \textbf{0.876 (0.022)} & \textbf{27.86 (1.02)} & 59.01 \\
    \midrule
    ViGU$_4$ & 15.07 & \underline{0.954 (0.009)} & 32.85 (1.05) & 26.06 & 0.868 (0.025) & 27.60 (1.03) & 63.43 \\
    ViGU$_4$-GAN & 15.07 & 0.949 (0.009) & 32.41 (1.02) & 22.44 & 0.841 (0.024) & 26.86 (0.91) & 87.03 \\
    ViGU$_2$ & 73.02 & \textbf{0.955 (0.009)} & \underline{32.95 (1.07)} & 22.73 & 0.872 (0.023) & 27.72 (1.02) & \underline{58.61} \\
    ViGU$_2$-GAN & 73.02 & \underline{0.954 (0.009)} & 32.88 (1.07) & \textbf{16.62} & \underline{0.873 (0.022)} & \underline{27.75 (1.00)} & \textbf{50.19} \\
    \bottomrule
    \end{tabular}%
    }
  \label{tab:comparison}%
\end{table}%
\begin{figure*}[t!]
    \centering
    \centerline{\includegraphics[width=5.6in]{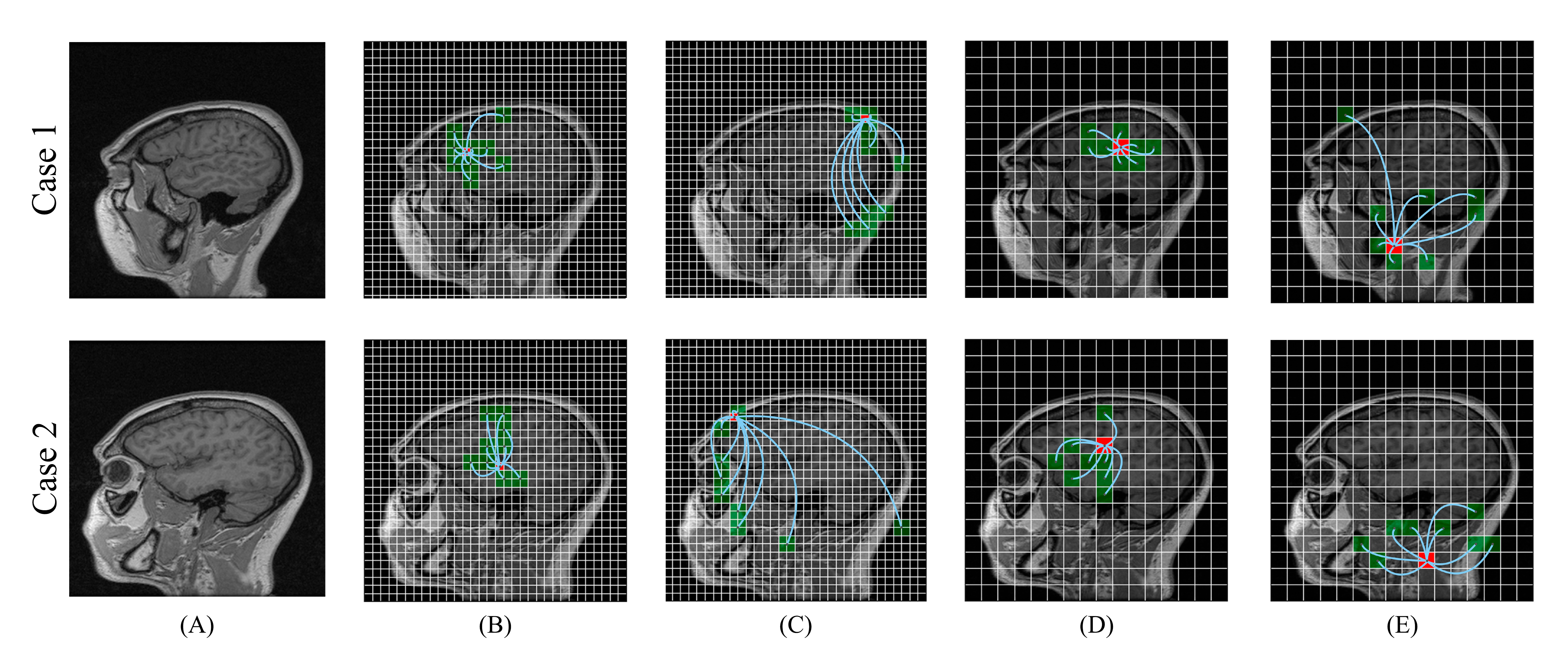}}
    \caption{Visualised graphs of the proposed ViGU. (A): the original MR images; (B-C): Graph connection from EnvB2; (D-E): Graph connection from EnvB3. A chosen node (red) and its first-order neighbours (green) are connected by edges (green line). In (B-C), $2 \times 2$ maximum pooling operation was applied for the neighbour nodes to reduce the computational cost.}
    \label{fig:FIG_Edge}
\end{figure*}
Table~\ref{tab:comparison} and Figure~\ref{fig:FIG_EXP_IMAGE_Comparison} show the quantitative results and visualised samples of the comparison experiments, respectively. The proposed ViGU and ViGU-GAN outperformed other CNN and GAN-based methods, and achieved comparable results compared to the Transformer-based method SwinMR, with only 1.9\% and 9.1\% MACs depending on the patch size. 

For the patch size setting, ViGU and ViGU-GAN with small patch sizes (larger patch resolution) tend to have better reconstructed results, whereas at the cost of larger MACs.

For the GAN-based variant ViGU-GAN, the utilisation of adversarial training mainly improves the perceptual experiments and reflects a better FID score. However, the proposed ViGU-GAN leads to an unstable training process (abnormal pool performance of ViGU$_4$-GAN using R10\% mask in Table~\ref{tab:comparison}), prolonged convergence time and enlarged GPU memory requirements. Further research and optimisation of GAN-based ViGU should be conducted.

\subsection{Visualised Graph \& Explainability}

Figure~\ref{fig:FIG_Edge} shows the visualised graph connection of the proposed ViGU, including reference MR images (A), and graph connection from EnvB2 (B-C) and EnvB3 (D-E). For better visualisation, we only display a chosen node (red) and its first-order neighbours (green) connected by an edge (green line). In Figure~\ref{fig:FIG_Edge} (B-C), $2\times2$ maximum pooling operation was applied for the neighbour nodes to reduce the computational cost, which led that the neighbour node area being bigger than the chosen node area.

The graph connection of the proposed ViGU model can provide an explainability of how the network recognises and extracts the feature of MR images.
Figures ~\ref{fig:FIG_Edge} (B) and (D) show that a node of brain tissue tends to have more neighbour nodes containing brain tissue, which proves that the network can be trained to gather the node with similar features and create the connection between them.
However, since there is no tag information added to the network, it is hard for the proposed ViGU to learn the accurate border of different without any supervision. Different anatomical structures with similar textures can also mislead the network. A node at the edge (border of the anatomical structures, not the edge in the graph) tends to have a neighbour node that is also at the edge, regardless of the anatomical structures (Figure~\ref{fig:FIG_Edge} (C) and (E)). 

\section{Discussion}
\label{sec:discussion}

This work has exploited how ViG works for MRI reconstruction, treating the MR images as graphs instead of conventional grid or sequence structure data. Using graph-based operation our proposed network can extract and process the feature more flexibly and efficiently since the irregular anatomical structures leads to redundancy and inflexibility using regular grid-based or sequence-based operations like CNN and transformers.
In addition, the proposed ViGU can learn a comprehensive understanding of the feature of MR images in latent non-Euclidean space, gathering and linking different parts with similar features globally.

In conclusion, we can envisage that our proposed ViGU and ViGU-GAN to be served as a UNet-based backbone for the graph-based MRI reconstruction, super-resolution and segmentation. For future work, segmentation information would be incorporated into the ViGU, guiding the network to build clinically-meaningful graphs, and improving the reconstruction performance while providing better explainability.

\vfill
\pagebreak

\section{Acknowledgements}
\label{sec:acknowledgments}
\sloppy
This study was supported in part by the ERC IMI (101005122), the H2020 (952172), the MRC (MC/PC/21013), the Royal Society (IEC\textbackslash NSFC\textbackslash211235), the NVIDIA Academic Hardware Grant Program, and the UKRI Future Leaders Fellowship (MR/V023799/1). CBS acknowledges support from the Philip Leverhulme Prize, the Royal Society Wolfson Fellowship, the EPSRC advanced career fellowship EP/V029428/1, EPSRC grants EP/S026045/1 and EP/T003553/1, EP/N014588/1, EP/T017961/1, the Wellcome Innovator Awards 215733/Z/19/Z and 221633/Z/20/Z, the European Union Horizon 2020 research and innovation programme under the Marie Skodowska-Curie grant agreement No. 777826 NoMADS, the Cantab Capital Institute for the Mathematics of Information and the Alan Turing Institute.

\bibliographystyle{IEEEbib}
\bibliography{strings,refs}

\end{document}